\documentclass[runningheads,11pt]{llncs}
\usepackage[T1]{fontenc}
\usepackage[a4paper, margin=3.8cm]{geometry}

\usepackage{amssymb,amsfonts,amsmath}
\usepackage{plain}
\usepackage{graphicx}
\begin{document}

%\numberwithin{equation}{section}
%\newtheorem{theorem}{Theorem}[section]
%\newtheorem{lemma}[theorem]{Lemma}
%\newtheorem{proposition}[theorem]{Proposition}
%\newtheorem{corollary}[theorem]{Corollary}
%\newtheorem{algorithm}[theorem]{Algorithm}
%\newtheorem{definition}[theorem]{Definition}
\spnewtheorem{algorithm}{Algorithm}{\bfseries}{\itshape}

\def\words#1{\quad\hbox{#1}\quad}
\def\wwords#1{\qquad\hbox{#1}\qquad}
\def\I{{\cal I}}
\def\J{{\cal J}}

\title{
	The Effects of Latency on a
	Progressive Second-Price Auction
}
\date{\today}
\author{Jordana Blazek, Eric Olson, Fredrick C. Harris, Jr.}
\institute{University of Nevada, Reno}
\maketitle
\begin{abstract}
The progressive second-price auction of Lazar and Semret 
is a decentralized mechanism for the allocation
and real-time pricing of a divisible resource.  
Our focus is on how delays in the receipt of bid messages,
asynchronous analysis by buyers of the market
and randomness in the initial bids affect the 
$\varepsilon$-Nash equilibria obtained by the
method of truthful $\varepsilon$-best reply.
We introduce an algorithm for finding minimal-revenue
equilibrium states and then show that
setting a reserve price just below clearing
stabilizes seller revenue while maintaining efficiency.
Utility is of primary interest
given the assumption of elastic demand.
Although some buyers experienced unpredictability in the value
and cost of their individual allocations, 
their respective utilities were predictable.
\end{abstract}

\section{Introduction}

Lazar and Semret~\cite{Lazar1999}
introduced a decentralized
second-price auction for allocation of network bandwidth in which 
buyers asynchronously update bid prices in a way that maximizes
their allocations while increasing their utility,
see also Semret, Liao, Campbell and Lazar~\cite{Semret2000}.
The novelty in this auction mechanism is a buyer's individual 
price valuation is revealed only locally when that same buyer 
updates their bid with a new quantity and corresponding marginal 
value.
Thus, the seller does not need the full price valuation curve of 
each buyer to allocate the resource in a way that maximizes 
the welfare or total value in the auction.

This progressive second-price auction was extended to the case
where prices were quantized by Jia and Caines in \cite{Jia2008}.
An opt-out 
selection method in the presence of multiple sellers was considered 
in \cite{Blocher2021}.
For markets consisting of many simultaneous auctions
a graph theoretic framework 
that characterizes influences resulting
from which buyers buy from which sellers in terms of a bipartite
graph is further explored in \cite{Blocher2025}.  

It is known provided $\varepsilon$ is large enough that 
Algorithm 1 in \cite{Lazar1999}
leads to bid updates that converge to an $\varepsilon$-Nash equilibium---an
equilibrium state where no buyer can individually increase 
individual their utility by more that $\varepsilon$ through 
changing their bid.
We call this the method of truthful $\varepsilon$-best reply.
Proposition 3 in \cite{Lazar1999} shows at such an 
equilibrium the auction is efficient, that is, the total value in 
the final allocation is within 
${\cal O}(\sqrt{\varepsilon\kappa})$ of optimal.
Here $\kappa$ is a constant related to the maximal
rate of diminishing returns among all buyers in the auction.

Consider three possible sources of randomness:  the initial bids of each buyer,
the asynchronous order in which buyers update their bids
and communication latency in the receipt of bid messages.
Perceptions of fairness are influenced by predictability of
outcomes; 
however, it is possible that randomness in the 
initialization and operation of the auction do not lead to 
significantly different equilibrium states.  
Our focus is on the degree the above sources of randomness 
influence both market aggregates and individual outcomes.

Second-price auctions, pioneered by Vickrey~\cite{vickrey1961}, allow the highest bidder to
pay the second-highest bid, incentivizing truthful bidding, as paying the second-highest
price minimizes the need for strategic manipulation. This auction type supports
fair pricing but faces challenges in dynamic and decentralized environments
where information and timing constraints can hinder optimal bidding,
see Brandt and Sandholm \cite{Brandt2008} and Maill\'e \cite{Maille2007}.

While aggregate quantities such as total value, utility and revenue
are of interest to the seller, an individual buyer is more concerned
about their individual costs, value and utility.
Assuming the buyers do not change, then it is desirable that
the allocation of resource between them also not change.
Our goal is to characterize the expected outcomes and
deviations in those outcomes for both the seller and individual
buyers that arise from the distributed asynchronous nature 
of the progressive second-price auction under a realistic
model of communication latency as described by
Arfeen, Pawlikowski, McNickle and 
Willig~\cite{Arfeen2019}, see
also Arshadi and Jahangir \cite{Arshadi2014}.

We begin our study by introducing an algorithm
to construct $\varepsilon$-Nash equilibria consisting of truthful bids
in which each buyer receives
the exact quantity requested and
consequently impose no externality on the other buyers in the auction.
As a result the second price for each buyer is zero and the
revenue received by the seller also zero.
Note that the existence of zero-revenue $\varepsilon$-Nash equilibria may
be inferred from Semret~\cite{Semret1999} who observed that
a buyer can obtain resource at zero cost by making a bid
with a zero per-unit price valuation.
On the other hand there are also $\varepsilon$-Nash equilibria
which generate revenue much closer to the total value in the auction.
Therefore, total revenue can vary significantly between different
$\varepsilon$-Nash equilibria of the same auction.
Since total utility is total value minus total
cost, it follows that
the total utility enjoyed by the buyers
can also vary significantly
between different equilibrium states.
%Note that Delenda, Maill\'e and Tuffin~\cite{Maille2004}
%demonstrate that the reserve price can be 
%optimized by simple numerical methods, allowing progressive second-price 
%auction to balance efficiency with revenue maximization.
Note that Delenda, Maill\'e and Tuffin~\cite{Maille2004}
balance revenue maximization with efficiency using an optimized
seller reserve price determined by simple numerical methods.

After setting the reserve price below the clearing price to reduce
the variability in revenue outcome for the seller without affecting
auction efficiency, we then proceed to
characterize the effects of latency and asynchronous bidding in the
progressive second-price auction.
Simulations indicate that the method of truthful $\varepsilon$-best
reply continues to converge even with high communications latencies.
Moreover, at the resulting equilibrium states,
although some buyers experience large deviations in the value
and cost of their respective allocations, the deviations in
their utilities are small.
Given the assumption of elastic demand, utility is of primary
importance and we find it remarkable how predictable this outcome was
when significant variability was present in the operation of the auction.

The structure of this paper is as follows:  Section~\ref{sec:prelim} 
introduces the formal setup, notation and buyer–seller valuation
framework. Section~\ref{sec:freelunch} analyzes zero-revenue equilibria, 
showing how truthful $\varepsilon$-best replies alternated with compromise
replies can lead to efficient 
but unprofitable outcomes, motivating the use of a reserve price.
Section~\ref{sec:latency} examines the role of communication latency 
and asynchronicity on equilibrium convergence and variability 
through stochastic modeling. Section~\ref{twinsec} details an
experiment consisting of many pairs of identical
twins in which one twin is lazy and evaluates the market more
slower and sends their bids with much greater latency.
Finally, Section~\ref{sec:conc} concludes 
with implications for decentralized market design and directions for 
future research.

\section{Preliminaries}\label{sec:prelim}

Consider a progressive second-price auction consisting of one seller and multiple buyers.
During the operation of the auction buyers place their bids
asynchronously and decentralization leads to communication latency
in the receipt of bid messages.
Thus, the outcome of the auction is affected by randomness
in the communication delays and bid ordering.
We also study what
effect the initial bids have on the outcome of the auction.

Suppose a quantity $Q$ of a divisible resource is to be allocated among
a fixed set of buyers. Let $\I$ be an index set such that $i\in\I$ 
represents a buyer able to participate in the auction. 
Each buyer has a price 
valuation $\theta_i(q)$ which identifies the value that buyer 
obtains upon receipt of a quantity $q$ of resource.  We suppose
the valuation increases in $q$ 
up to a maximum quantity $\bar q_i$ and is concave to 
reflect diminishing returns.
Assuming $\theta_i$ is differentiable, it follows that $\theta_i'$
is decreasing and the 
greatest marginal value buyer~$i$ ever places on the resource 
is given by $\bar p_i=\theta_i'(0)$.

Intuitively, $\theta_i'(q)$ represents the
value of the next unit of resource after $q$ units have been obtained.
Take $\theta_i$ 
as in \cite{Lazar1999} 
to be quadratic of the form
\begin{equation}\label{valuation}
\theta_i(z)=\begin{cases}
	(1- {1\over 2} z/\bar q_i)z\bar p_i &\hbox{for $z<\bar q_i$}\cr
	{1\over 2}\bar q_i \bar p_i& \hbox{otherwise.}
\end{cases}
\end{equation}
where buyer demand $\bar q_i$ is sampled uniformly over
$[50, 100]$ and the maximal marginal valuation $\bar p_i$
is uniform on $[10, 20]$.
Note that the resulting decrease in marginal value
is bounded uniformly in $i$ both above and below.
Unless otherwise mentioned we consider $100$ buyers that participate
in the auction and keep their respective price valuations $\theta_i$
fixed as
well as the amount of resource $Q=1000$ available in the auction.

\begin{remark}\label{scarcity}
Since $\bar q_i\ge 50$ for each buyer, the total quantity
of resource valued by $100$ buyers is at least $5000$.  Therefore
$Q=1000$ is guaranteed to be a condition of scarcity in which at least
one buyer has the potential to increase the value of their allocation.
\end{remark}

Now consider the bids $(q_i,p_i)\in [0,\bar q_i]\times [0,\infty)$
from all buyers $i\in{\cal I}$ able to participate in the auction.
Here $q_i$ is the quantity requested and $p_i$ the amount the 
buyer would be willing to spend to obtain an additional unit of 
resource, that is, the marginal value at $q_i$.
Thus, a truthful bid always satisfies $p_i=\theta_i(q_i)$.  
We emphasize that the full valuation curve $\theta_i$ for each buyer
is not made available to the seller but revealed only locally
though the marginal value $p_i$ at the quantity $q_i$ being 
bid on.

Denote the bid $(q_i,p_i)$ submitted by buyer $i$
as $(i,q_i,p_i)$.  
Then the set of all bids in the auction is
$
	s=\big\{\, (i,q_i,p_i): i\in\I\,\big\}.
$
We let the bid $(0,Q,P)$
represent the reserve price set by the seller,
never update this bid and suppose $0\in\I$.
In this special case $\theta_0(a)=Pz$ is linear and $\theta_0'(z)=P$ constant.
For convenience denote $\I_0=\I\setminus\{0\}.$

If $k\ne i$ then buyer $k$ is in competition with buyer $i$ and
the opposing bids against which buyer $i$ must bid are
$s_{-i}=\big\{\, (k,q_k,p_k)\in s : k\ne i\,\big\}$.

At a marginal price of $y$ the resource 
available to buyer $i$ is
\begin{equation}
	Q_i(y,s_{-i})=\max\{Q-z,0\}
\end{equation}
where
$
	z=\textstyle\sum \big\{\,
    q_k : (k,q_k,p_k)\in s_{-i}\hbox{ and }p_k>y\,
    \big\}.
$
Note that if $y<P$ the reserve price takes effect,
the bid $(0,Q,P)$ implies $q_0=Q$ is an addend in
$z$ and consequently $Q_i(y,s_{-i})=0$.

Conversely, obtaining at least a quantity $z$ of resource from the auction
requires a bid with marginal price
\begin{equation}
    P_i(z,s_{-i})=\inf\big\{\,
        y\ge 0:Q_i(y,s_{-i})\ge z\,\}.
\end{equation}

Practically speaking, the progressive second-price auction aims
for an $\varepsilon$-Nash equilibrium in which 
an individual buyer's utility can not be increased by more than 
$\varepsilon$.
At such equilibria bid prices receving a positive allocation of resource
will be close but not in general the same.
Note, however, that if prices are quantized---for example in dollars
and cents---then ties become more likely, especially when $\varepsilon$ is small.
For the simulations in this paper we do not quantize prices and
take $\varepsilon=5$ throughout.
Even so, a tie breaking condition appears useful.
Following Jia and Caines \cite{Jia2008} we allocate bids whose
prices are tied proportionally.

Thus, the allocation to buyer $i$ is
\begin{equation}\label{alloc}
    a_i(s)=(q_i/z)\min\big\{q_i,Q_i(p_i,s_{-i})\big\}
\end{equation}
where $z=1$ if $q_i=0$ and otherwise
$$
    z=\textstyle\sum\big\{\, q_k : (k,q_k,p_k)\in s
    \hbox{ and }p_k=p_i\,\big\}.
$$

Since the cost $c_i(s)$ to buyer $i$ for participating in a
second-price
auction is the loss incurred by the other buyers due to
that participation, changes in the allocations can be used to
compute costs.  In particular,
\begin{equation}
	c_i(s)
		=\sum_{k\in\I\setminus\{i\}}
			p_k \big(a_k(s_{-i})-a_k(s)\big).
\end{equation}

\begin{remark}\label{instante}
As our second-price auction is progressive,
the bids $s_{-i}$ reflect
the historical influence of buyer $i$ leading up to the present time.
Thus, $c_i(s)$ represents the
externality obtained by omitting buyer~$i$
when determining the allocation and not the true externality
that would have resulted if
buyer $i$ were never part of the auction.
This difference between the true and instantaneous externality
appears central to the
zero revenue $\varepsilon$-Nash equilibrium in the next section.
\end{remark}

Unlike \cite{Lazar1999}, \cite{Jia2008} and \cite{Blocher2021} we 
do not consider each buyer to be further constrained 
by a budget $b_i$ that bounds the cost they are willing to
incur for their resource allocation.  This is for simplicity
and to avoid situations where the second price changes in
such a way that a previous bid needs to be updated to stay
under budget.

Crucial to the progressive second price
auction is the maximum allocation available in a market on 
the price valuation curve of buyer $i$.
Again following \cite{Lazar1999} we consider
the bid update rule given by 
\begin{definition}\label{ebest}
The truthful $\varepsilon$-best reply is defined as follows.  Let
\begin{equation}\label{gidef} 
	G_i(s_{-i})
	=\big\{\, z\in [0,\bar q_i]:z \le Q_i(\theta_i'(z),s_{-i})\,\big\}.
\end{equation}
The $\varepsilon$-best reply to the bids $s_{-i}$ is defined as 
$$
	(v_i,\theta_i'(v_i))
	\wwords{where} v_i=\sup G_i(s_{-i})-\varepsilon/\theta_i'(0).
$$
\end{definition}

The above bid is truthful since it lies on the graph of the marginal value.
Taking $v_i$ slightly less than the supremum ensures $v_i\in G_i(s_{-i})$ and
increases the bid price $\theta_i'(v_i)$ due to decreasing marginal value (except
in the case $i=0$).
The term $\varepsilon/\theta_i'(0)$ ensures the bid $(v_i,\theta_i'(v_i))$
is within $\varepsilon$ of the best bid.
In other words, the truthfull $\varepsilon$-best reply of
Definition~\ref{ebest} provides a truthful bid that cannot
be improved by more than $\varepsilon$ while holding the opposing
bids constant.

%A proof of the above fact depends on two different ways of
%viewing the cost---the exclusion-compensation theorem---which
%we state here as
%\begin{theorem}\label{compensation}
%In a progressive second price auction 
%under the above assumptions we have
%\begin{equation}\label{eq:cost}
%	c_i(s)=
%	\int_0^{a_i(s)}P_i(z,s_{-i}) dz.
%\end{equation}
%\end{theorem}
%\noindent

%Theorem~\ref{compensation} appears as equation (8) 
%in \cite{Lazar1999}
%along with the observation that repeatedly allowing each buyer
%to update their $\varepsilon$-best replies converges to 
%an $\varepsilon$-Nash equilibrium provided $\varepsilon$ is large
%enough.

\section{Zero-Revenue Equilibria}\label{sec:freelunch}

Before studying how communication latency and asychronous bidding affect the outcomes of the progressive
second-price auction, we first address variations in outcomes
that result from Remark~\ref{instante} on the computation of cost.
This section examines zero-revenue equilibria. 
It demonstrates how such equilibria emerge under the standard mechanism
and explains why a reserve price is required to ensure positive seller revenue.

To construct zero-revenue equilibria 
we modify the method of truthful $\varepsilon$-best
reply given as Algorithm 1 in \cite{Lazar1999} to alternate between the original bidding strategy and a 
compromise bid.  Simulations indicate this modified algorithm still 
converges to an $\varepsilon$-Nash equilibrium, but in this case one
in which each buyer obtains the exact allocation they asked for.

While the $\varepsilon$-best reply chooses a near optimal bid without
regard for a buyers previous bid,
a compromise bid is based on the requested and
and allocated resource from the previous bid.  Namely, we have
\begin{definition}\label{compromise}
Suppose the bid $(q_i,p_i)$ from buyer $i$ receives an allocation 
of $a_i(s)$.  The truthful compromise reply is defined as
$$(v_i,\theta'_i(v_i))\wwords{where}
	v_i=(a_i(s)+q_i)/2.
$$
\end{definition}

While compromise bids generally do not lead to an 
$\varepsilon$-Nash equilibrium on their own,
interesting behavior happens when bids are 
alternated between the $\varepsilon$-best reply 
of Definition~\ref{ebest} and the compromise reply of 
Definition~\ref{compromise}.
The compromise bids reduce the externality 
while the $\varepsilon$-best replies
advance towards an $\varepsilon$-Nash equilibrium.
%The result is a $\varepsilon$-Nash equilibrium in
%which each buyer receives exactly what they requested
%and consequently impose no externality on any of the
%other buyers.  The second price is zero and the
%seller receives zero revenue.

First recall the method of truthful $\varepsilon$-best reply from
\cite{Semret1999} as
\begin{algorithm}\label{ebestalg}
The method of truthful $\varepsilon$-best reply is
\begin{itemize}
\item[1.] Each buyer evaluates the $\varepsilon$-best reply (Definition~\ref{ebest}) and updates bids when utility increases by at least $\varepsilon$.
\smallskip
\item[2.] Repeat until no additional $\varepsilon$-best replies occur.
\end{itemize}
\end{algorithm}

\begin{remark}
At first it seems plausible that only a subset of 
$\varepsilon$-Nash equilibria might be obtained through 
the method of truthful $\varepsilon$-best reply and 
that the zero-revenue case might not be among them.
However, if the initial bids start at a particular 
equilibrium state, then the $\varepsilon$-best 
reply will remain at that equilibrium.
Thus, the method of $\varepsilon$-best reply can
terminate at any equilibrium state simply by starting at
that equilibrium.
\end{remark}

\begin{remark} The $\varepsilon$-best reply for buyer $i$ is constructed
so their allocation $a_i(s)=q_i$; however, increasing the utility of
buyer $i$ will generally cause some of the opposing buyers to lose their
part of their allocations.  Thus, after an $\varepsilon$-best reply
there may be $j\ne i$ such that
$a_j(s)<q_j$.  This reflects an externality imposed on buyer $j$.
If buyer $j$ makes a compromise bid in reply, this reduces the
externality imposed by buyer $i$ on buyer $j$ while at the same
time not reducing the allocation of buyer $j$. In particular, switching
between $\varepsilon$-best and compromise bids tends to an
$\varepsilon$-Nash equilibrium in which no other buyer imposes
any externality on another buyer.
\end{remark}

Our modified algorithm may now be stated as
\begin{algorithm}\label{compalg}
The alternating $\varepsilon$-best with compromise bid method is
\begin{itemize}
\item[1.] Each buyer evaluates the $\varepsilon$-best reply (Definition~\ref{ebest}) and updates bids when utility increases by at least $\varepsilon$.
\smallskip
\item[2.] Each buyer submits a compromise bid (Definition~\ref{compromise}) independent of immediate utility gain.
\smallskip
\item[3.] Repeat until changes in compromise bids fall below tolerance and
no additional $\varepsilon$-best replies occur.
\end{itemize}
\end{algorithm}

Note that compromise bids are made
whether or not they increase utility.
We do not comment on the rationality of this method of bidding but 
instead view Algorithm~\ref{compalg} as a technique to obtain 
zero-revenue $\varepsilon$-Nash equilibria.
In turn such equilibria
demonstrate the need for a reserve price even though the
allocations are near value maximizing.

To demonstrate the convergence of the alternating $\varepsilon$-best
with compromise bid method
to a zero-revenue $\varepsilon$-Nash equilibria, consider a simplified 
version of the progressive second-price auction with no 
communication latency or asynchronous bidding.  In this case 
buyers bid round-robin and alternate between bid strategies.  The
only source of randomness comes from the initial bids.  We
illustrate the effects of the reserve price by choosing different
values of $P$ ranging from $0$ to $16$.

Given an equilibrium state $s$ obtained from Algorithm \ref{compalg}
we are interested in the aggregate quantities of revenue,
total value and total utility.  These are given, respectively, by
$$
    S[c]=\sum_{i\in\I_0} c_i(s),\quad
    S[v]=\sum_{i\in\I_0} \theta_i(a_i(s))\words{and}
    S[u]=S[v]-S[c].
$$
To further understand the effects of the reserve price, we also
compute an average bid price as
$$
    E[p] = \frac{1}{S[a]}
        \sum_{i\in\I_0} a_i(s) p_i
\wwords{where}
    S[a] = \sum_{i\in\I_0} a_i(s).
$$

Study how the above aggregate quantities depend on the initial bids
by averaging them over the $\varepsilon$-Nash equilibira corresponding to
an ensemble of 100 randomly chosen initial bids.
Specifically,
consider a random set of initial bids of the form
$\{\,(i,q_i,\theta_i'(q_i)): i\in\I\,\}$ where the
$q_i$ are independent and uniformly distributed over $[0,\bar q_i]$
for $i\ne 0$ and $q_0=Q$.  Let $\Omega$ be the underlying uniform
sample space.  Given
$\omega\in\Omega$ define $s_w$ to be the $\varepsilon$-Nash
equilibrium obtained from
Algorithm~\ref{compalg} 
starting at the initial bid corresponding to $\omega$.

Now let ${\cal E}\subseteq\Omega$ be an ensemble of $100$ independent
realizations of the initial bids.
Suppose $X$ is either revenue, total value, total utility or bid price.
The ensemble
average and sample variance of $X$ is given by
\begin{equation}\label{xensemble}
    \langle X\rangle = {1\over |{\cal E}|}\sum_{\omega\in{\cal E}}X_w
    \words{and}
    V[\langle X\rangle]={1\over |{\cal E}|-1}
        \sum_{\omega\in{\cal E}} \big(X_w-\langle X\rangle\big)^2.
\end{equation}
Here $X_w$ indicates $X$ measured at the $\varepsilon$-Nash
equilibrium given by $s_w$.  For example, if $X=S[c]$ then the ensemble-averaged
revenue would be
$$
    \langle S[c]\rangle = {1\over |{\cal E}|}\sum_{\omega\in{\cal E}} 
    \sum_{i\in\I_0} c_i(s_w).
$$

\begin{table}[h]
    \centerline{\begin{minipage}[b]{0.75\textwidth}
    \caption{\label{reserve}%
    The effects of reserve price on the bid price,
    total value, utility and revenue in the
    $\varepsilon$-Nash equilibria
    obtained from Algorithm \ref{compalg} averaged over 100 different random initial bids.
    Except for the revenue corresponding to a zero reserve price,
    the standard deviations---not shown---were
    less than 1 percent of the averages.
    }
    \end{minipage}}
\medskip
\centerline{
    \begin{tabular}{l r r r r r}
    \hline
    \noalign{\smallskip}
    Reserve Price&0&6&12&14&16\\
    \noalign{\smallskip} 
    \hline
    \noalign{\smallskip}
    Bid Price&\quad 13.4024&\quad 13.4022&\quad 13.3745&\quad 14.124&\quad 16.1289\\
    Total Value&\quad 15544.5&\quad 15544.6&\quad 15536.8&\quad 12623.1&\quad 5784.98\\
    Total Utility&15544.5&9544.56&3536.76&1602.27&441.175\\
    Total Revenue&$10^{-13}$&6000&12000&11020.8&5343.8\\
    \noalign{\smallskip} 
    \hline
    \end{tabular}}
\end{table}

Table~\ref{reserve} reports ensemble averages obtained through
simulation of Algorithm~\ref{compalg} over an ensemble of $100$ different
initial bids.  The standard deviations $V[\langle X\rangle]^{1/2}$ were consistently
less than $1$ percent of the average except for the zero-revenue case corresponding
to the reserve price of $P=0$ where both the average and deviation were numerically
equal to zero.
Note that for $P\le 12$ the
total revenue is equal to $QP$ but when $P\ge 14$ the total revenue decreases.
This is because as that point the seller starts buying back the resource which
does not contribute to revenue.

\begin{remark}
Since $\theta_i'(z)$ is decreasing, it is invertible.  Define
$$
    f_i(y)=\begin{cases}
        (\theta_i')^{-1}(y) & \hbox{for $y\in [0,\bar p_i]$}\cr
        0 & \hbox{otherwise.}
    \end{cases}
$$
If the reserve price $P$ is such that
$\sum_{i\in\I_0} f_i(P)<Q$
then seller buyback is guaranteed.
A fundamental premise of
the progressive second-price auction
is that the seller does not know the full valuation of
each buyer; therefore,
we do not explore this condition
further in this paper.
\end{remark}

Given $w\in\Omega$ observe that the revenue-minimizing equilibrium $s_w$ is
contained in a neighborhood of truthful bids which are nearly revenue
minimizing and also $\varepsilon$-Nash equilibria.  It follows that if
the initial bids are chosen randomly, then there is a positive probability
that those bids are already at an $\varepsilon$-Nash equilibium that is
nearly revenue minimizing.  

\section{Latency and Asynchronicity}\label{sec:latency}

This section focuses on the
variability caused by asynchronous bidding and communication latency
on the outcomes of the $\varepsilon$-Nash
equilibrium states reached using Algorithm~\ref{ebestalg} the method of 
truthful $\varepsilon$-best reply.
Set $P=12$ near the largest value such that all resource is allocated.
This maintains efficiency while reducing the variability in revenue
between equilibrium states.  Also fix the initial bids
made by the buyers.

We use a renewal processes to model both communication latency 
and asynchronicity in bid updates.  Namely, we employ a sequence
of independent Weibull-distributed random variables with probability
density
$$
	{\rm pdf}(x)=\frac{\beta}{\lambda} \Big(\frac{x}{\lambda}\Big)^{\beta-1} e^{-(x/\lambda)^\beta}
$$
where $\beta$ is the shape and $\lambda$ the scale.
The parameter $\beta$ characterizes the delay regime: $\beta < 1$ 
corresponds to a heavy-tailed, light-traffic latency distribution 
(bursty communication), whereas $\beta > 1$ represents a controlled, 
scheduled update process with more predictable timing.
Inspired by \cite{Arfeen2019} and \cite{Arshadi2014} we model
the interarrival times of bid messages using 
$\lambda_c=1.0$ and $\beta_c=0.75$
with a translation of $\delta_c=0.1$ seconds.
The decreasing hazard rate 
represents a bid sent under 
conditions of light traffic where most messages arrive with 
minimum latency but when lost experience exponential backoff 
of retransmission times.

\begin{figure}[h!]
    \centerline{\begin{minipage}[b]{0.75\textwidth}
    \caption{\label{weibull}%
	Comparison of the probability density functions governing
	the time between evaluation of bids and the communication
	latency to transmit a bid to the auction.}
    \end{minipage}}
    \centerline{
        \includegraphics[height=0.4\textwidth]{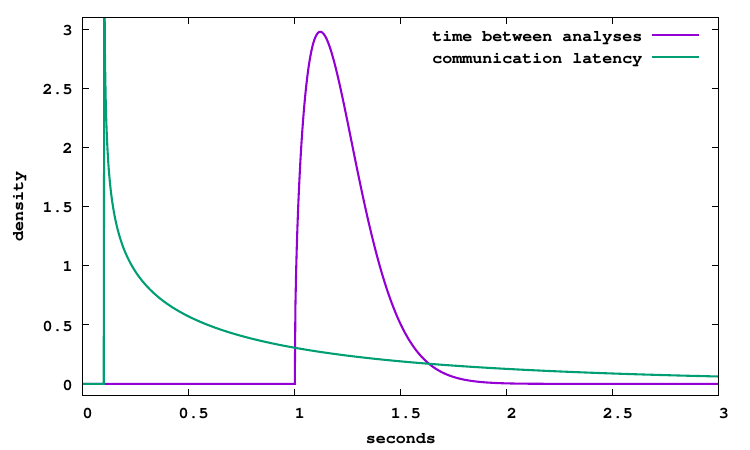}
    }
\end{figure}

As the auction progresses the $\varepsilon$-best reply is
evaluated by each buyer over
intervals characterized by 
a Weibull distribution 
with $\lambda_e=0.25$ and $\beta_e=1.5$ 
translated by $\delta_e=1.0$ seconds.
The increasing hazard rate represents a
controlled scheduling of market analysis that, in part, 
results from the assumption that each bid incurs a 
cost of $\varepsilon$ that needs to be amortized before 
making the next bid.

\begin{remark}
For each fixed buyer the intervals between
market analyses follow a Weibull distribution; however,
actually sending a message to update a bid is only
performed when the utility can be
increased by at least $\varepsilon$.  Moreover, as the state
of the auction gets closer to an equilibrium, the rate at
which $\varepsilon$-best replies lead to a bid update
appears to slow down.
\end{remark}

Let $\Delta^c_i$ represent
the communication delay (time from bid placement to activation) and
$\Delta^e_i$ be the evaluation interarrival delay
(time between attempted rebids).  These random variables are
independent and distributed according to
$$
\Delta^c_i \sim \delta_c+\mathrm{Weibull}(\lambda_c, \beta_c)
\quad \text{and} \quad
\Delta^e_i \sim \delta_e+\mathrm{Weibull}(\lambda_e, \beta_e).
$$
Thus, a bid $(q_i,p_i)$
computed at time $t$ is observed at time $t + \Delta^c_i$.
Similarly, a buyer who
analyzes the current state of the market at time $t$ will
again compute their $\varepsilon$-best reply at time $t+\Delta^e_i$.
The simulation operates as a discrete-time event system. Events are scheduled 
and processed in a priority queue, advancing the simulation clock $t$ to
the next event. 

Figure~\ref{weibull} depicts the distributions of the time between
bids and the communication latency.
Even though the expected time between bid updates is much greater than
the expected latency, the heavy tail corresponding to the shape
parameter $\beta_c=0.75$ implies there is a chance---due to network lag---that a new
bid update might be considered before the previous bid has been
received by the auction. 

To characterize the effects of latency and asynchronicity on the
outcomes of the progressive second-price auction we consider an
ensemble ${\cal E}$ of $100$ realizations for the random processes
given by $\Delta_i^c$ and $\Delta_i^e$.  Note that the ensemble
used in Section~\ref{sec:freelunch} was over random initial bids.
In this section we hold the initial bids fixed.
Now, for $\omega\in{\cal E}$ let $s_w$ denote the
equilibrium state obtained from the Algorithm~\ref{ebestalg}
subject to the communication
delays and market evaluations specified by $\omega$.

\begin{figure}[h!]
    \centerline{\begin{minipage}[b]{0.75\textwidth}
    \caption{\label{costvalueplot}%
	Left shows that changing the scale
    $\lambda_{\rm c}$ of the communication latency
    has minimal effect on the ensemble-averaged price
    and total utility received by all buyers.
    Right shows the average value and cost for
    each individual buyer when $\lambda_c=1$.}
    \end{minipage}}
    \medskip
    \centerline{
        \includegraphics[height=0.4\textwidth]{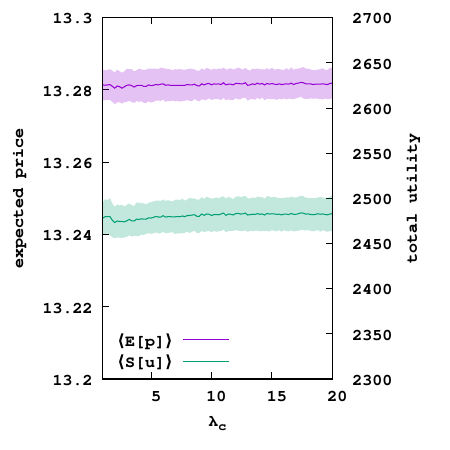}
        \includegraphics[height=0.4\textwidth]{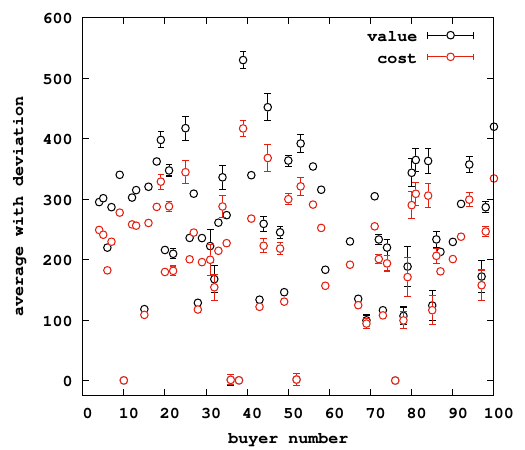}
    }
\end{figure}

Figure~\ref{costvalueplot} on the left depicts the ensemble
averages of the bid prices and total utility as a function of
the scale of the communications latency with $\lambda_c$ 
ranging from the default value of $1.0$ shown in Figure~\ref{weibull}
up to $20$.  The shadows illustrate the standard
deviations of the ensembles given by $V[\langle E[p]\rangle]^{1/2}$
and $V[\langle S[u]\rangle]^{1/2}$.
The deviations are small while the bid prices and total utility
are horizontal to within the errors.

%Note that the variance of the ensemble average $V[\langle E[p]\rangle]$ is
%different than ensemble average of the variance, which instead is given by
%$$
%\langle V[p]\rangle = {1\over |{\cal E}|}\sum_{\omega\in{\cal E}}
%\bigg(
%\frac{1}{S[a]}\sum_{i\in\I_0} a_i(s) (p_i - E)^2\bigg)\bigg|_{s=s_w}.
%$$

On the right Figure~\ref{costvalueplot} depicts the predictability
of individual outcomes for the $62$ buyers who received allocations in the
market.  The remaining~$28$ buyers consistently did not receive
allocations and are not shown.  
%No resource was purchased by the seller.
We plot the values and costs for each buyer given by respectively 
setting $X$ equal $v_i$
and $c_i$ in \eqref{xensemble}.
%$$
%    \langle v_i \rangle={1\over|{\cal E}|}\sum_{\omega\in{\cal E}} v_i(s_w)
%    \wwords{and}
%    \langle c_i \rangle={1\over|{\cal E}|}\sum_{\omega\in{\cal E}} c_i(s_w)
%$$
%and the standard deviations in these averages.
Some buyers experienced much higher deviations in outcomes compared to
others.  This also happened to the same buyers over different ensembles.
Similar results affecting
different individuals were observed for other populations of buyers.

Since demand is perfectly elastic, then
buyer utility $u_i=v_i-c_i$ is arguably more important than either
the value or cost on their own.
The deviation $V[\langle u_i\rangle]^{1/2}$ was computed for the
ensemble shown in Figure~\ref{costvalueplot} and found to be 
uniformly less than $1.3$ for all buyers while deviations
in cost and value were more than $25$ for some buyers.

\section{Different Latencies}\label{twinsec}
This section studies whether buyers who compute their $\varepsilon$-best
reply more frequently and experience less latency in their bid messages have
any advantage over buyers who analyze the market less frequently and whose
bid messages suffer greater latencies.
Consider a population of 100 buyers with valuation curves such that
$$
    \theta_{i+50}=\theta_i\wwords{for}i=1,2,\ldots,50.
$$
The first 50 buyers are identical twins of the last 50 buyers with
one difference:  The last 50 buyers are lazy and evaluate the market
less frequently and experience more latency in
their bid messages.

Specifically, the first 50 buyers keep the same bid evaluation
frequency and latency as in Table~\ref{weibull} while the delay and
scale parameters $\delta_c$, $\delta_e$, $\lambda_c$ and $\lambda_e$
for the last 50 buyers are multiplied by a factor of 17.  This leads to
an auction with two time scales:  One set by the industrious buyers
and the other by the lazy buyers.

\begin{figure}[h!]
    \centerline{\begin{minipage}[b]{0.75\textwidth}
    \caption{\label{lazy}
	The outcomes for lazy buyers who evaluate the market 17 times
    less frequently and experience 17 times the latency in their
    bid messages compared to an equal number of industrious buyers with identical
    valuations.}
    \end{minipage}}
    \medskip
    \centerline{
        \includegraphics[height=0.4\textwidth]{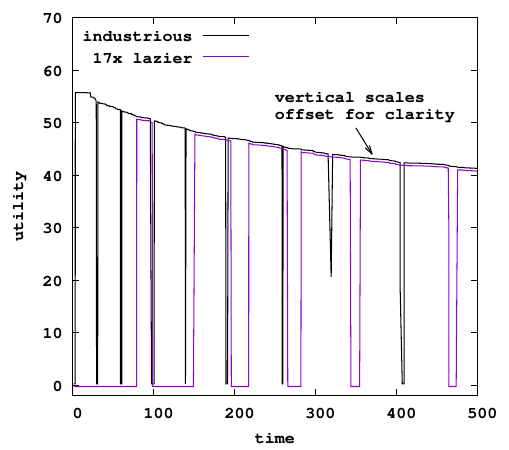}
        \includegraphics[height=0.4\textwidth]{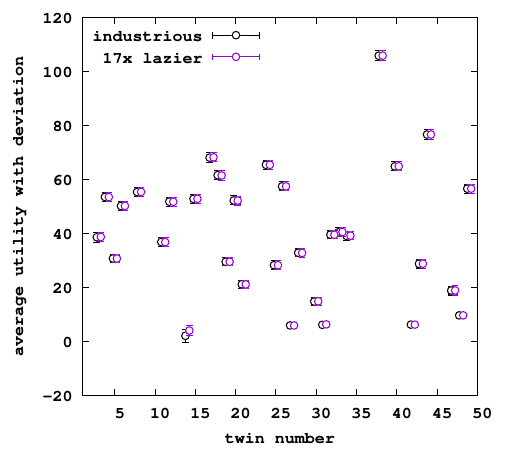}
    }
\end{figure}

Figure \ref{lazy} depicts the outcomes of the first 50 buyers compared
their identical but lazy twins, the last 50 buyers.  
The graph on
the left shows the transient part of the utility received over time
for representative pair of buyer twins.  After making an
$\varepsilon$-best reply, either twin obtains the same utility from
the market as the other.  However, the difference in the time scales
allows the industrious twin to maintain non-zero utility for a larger
percentage of the time.  As the equilibrium state is reached---not shown---the
rate of bid updates slows down so much that the fact that one twin
is 17 times slower than the other no longer matters.

In the end, both
twins obtain essentially the same utility at the equilibrium state.
Figure \ref{lazy} on the right illustrates the statistics for the same twin pairs
taken over an ensemble ${\cal E}$ of size 100.  The outcomes
in terms of individually received utility are
indistinguishable between the industrious and lazy twin.  
There is also very little deviation in outcomes due to the asynchronous nature
of the
individual market evaluations and the communication latencies.
This is consistent with the uniformly small deviations in utility 
observed in the previous section.

\section{Conclusions}\label{sec:conc}

We have introduced  a method to 
construct $\varepsilon$-Nash equilibria consisting of truthful
bids in which each buyer imposes zero externality on the other
buyers and
demonstrated through simulation that this algorithm converges.
Although it is known that the method of truthful $\varepsilon$-best
replies
may fail to converge unless $\varepsilon$ is large enough, it is possible
the introduction of compromise bids removes the condition on $\varepsilon$.
A theoretical proof of such a possibility seems out of reach; however, a more
modest result is planned for future work.  Namely, if
one assumes Algorithm~\ref{ebestalg} converges, then we conjecture 
Algorithm~\ref{compalg} preserves that convergence.

Including a reserve price at which the seller
purchases their own resource prevents
the zero-revenue equilibrium states found by Algorithm~\ref{compalg}
and reduces the variability in revenue outcomes
without reducing efficiency, provided the reserve is not set too high.
Although it is possible that a revenue-minimizing
equilibrium occurs from the random initialization of bids, an
interesting avenue for future study is whether compromise bids
made by a subset of buyers---perhaps organized through social media---would
be sufficient to reach a revenue-minimizing equilibrium in the presence of
other buyers who continue to bid only $\varepsilon$-best replies.

We have considered renewal processes governed by Weibull
distributions to model communication latency and the delay between
market evaluation times.
While the individual outcomes of value and cost had large deviations
for some buyers, deviations in utility were uniformly small.
This is a desirable property since utility is the important outcome
for the case of elastic demand considered in this auction model.

Communication delays imply the truthful $\varepsilon$-best replies
while still truthful might no longer be $\varepsilon$-best by the
time they reach the seller.  Therefore, we were surprised that 
Algorithm~\ref{ebestalg} continued to converge no matter how much
we increased the latency.
Moreover,
given the dramatically different timescales for the random processes
employed in the identical twin simulations,
even more remarkable was that the individual utility and bid prices
at the $\varepsilon$-Nash equilibria had so little dependency on the
random elements in the market mechanism.
These are desirable properties for any
distributed asynchronous method of finding optima.

In accordance with Algorithm~\ref{ebestalg} our simulations
send a new bid---whether the previous bid has been received or not---only
if the new bid is $\varepsilon$-better than the previously sent bid.
Because of the decreasing hazard rate in the communications network,
a previously sent bid which has not been activated yet is even less
likely to be activated as time goes on.  For this reason it could be advantageous
for a buyer to send a new bid that does not improve on a previous
bid simply because the previous bid has lagged in the network.  
The authors consider such a modification
to Algorithm~\ref{ebestalg} and
similar modifications based on amortization of bid cost over time
to be interesting directions for future research.

\bibliographystyle{plain}

\end{document}